\documentclass[fleqn,usenatbib]{mnras}

\usepackage{newtxtext,newtxmath}
\usepackage[T1]{fontenc}

\DeclareRobustCommand{\VAN}[3]{#2}
\let\VANthebibliography\thebibliography
\def\thebibliography{\DeclareRobustCommand{\VAN}[3]{##3}\VANthebibliography}

\usepackage{graphicx}	% Including figure files
\usepackage{amsmath}	% Advanced maths commands

\usepackage{ulem}

\newcommand{\dd}{\mathrm d}
\renewcommand{\d}{\partial}

\renewcommand{\(}{\left(}
\renewcommand{\)}{\right)}
\renewcommand{\[}{\left[}
\renewcommand{\]}{\right]}
\renewcommand{\v}[1]{\boldsymbol{#1}}

\title[Can superbubbles accelerate UHE protons?]{Can superbubbles accelerate ultra-high energy protons?}

\author[T. Vieu et al.]{
T. Vieu,$^{1}$\thanks{E-mail: thibault@mpi-hd.mpg.de}
B. Reville,$^{1}$
F. Aharonian$^{1,2}$
\\
$^{1}$Max-Planck-Institut f\"ur Kernphysik, Saupfercheckweg 1, D-69117 Heidelberg, Germany\\
$^{2}$Dublin Institute for Advanced Studies, 31 Fitzwilliam Place, Dublin 2, Ireland
}

\date{Accepted XXX. Received YYY; in original form ZZZ}

\pubyear{2022}

\begin{document}
\label{firstpage}
\pagerange{\pageref{firstpage}--\pageref{lastpage}}
\maketitle

\begin{abstract}
We critically assess limits on the maximum energy of protons accelerated within superbubbles around massive stellar clusters, considering a number of different scenarios. In particular, we derive under which circumstances acceleration of protons above peta-electronvolt (PeV) energies can be expected. While the external forward shock of the superbubble may account for acceleration of particles up to 100~TeV, internal primary shocks such as supernova remnants expanding in the low density medium or the collective wind termination shock which forms around a young compact cluster provide more favourable channels to accelerate protons up to $1$ PeV, and possibly beyond. Under reasonable conditions, clustered supernovae launching powerful shocks into the magnetised wind of a young and compact massive star cluster are found to be the most promising systems to accelerate protons above 10~PeV. On the other hand, stochastic re-acceleration in the strongly turbulent plasma is found to be much less effective than claimed in previous works, with a maximum proton energy of at most a few hundred TeV. 
%something about pulsars here...?
\end{abstract}

\begin{keywords}
cosmic rays -- acceleration of particles -- open clusters and associations: general -- shock waves
\end{keywords}

%%%%%%%%%%%%%%%%%%%%%%%%%%%%%%%%%%%%%%%%%%%%%%%%%%

%%%%%%%%%%%%%%%%% BODY OF PAPER %%%%%%%%%%%%%%%%%%

\section{Introduction}
The search for Galactic cosmic-ray (CR) sources able to accelerate above PeV energies has been ongoing for decades. The importance of this energy range was established early on, as it presented an obstacle to CR origin theories via diffusive shock acceleration (DSA) at supernova remnant (SNR) shocks \citep{lagage1983,Heavens,axford1994}.
For standard ISM magnetic field values, and assuming the most optimistic conditions for transport of CRs near the shock (i.e. the Bohm diffusion limit) 
%Assuming quasi-linear saturation of resonantly excited MHD fluctuations, i.e. $\left\langle\delta B^2\right\rangle \lesssim B_0^2$, 
the maximum energy for typical SNRs was shown to fall close to 100~TeV \citep[e.g.][]{cesarsky1981ICRC,lagage1983c}. Such low maximum energies could not account for the knee feature in the CR spectrum at a few PeV, let alone the steeper second Galactic population extending above the knee to $\sim 10^{18}$~eV energies.  
%which could not explain the transition between the population of Galactic and extragalactic CR in the range $10^{15} - 10^{18}$~eV. 
Alternative scenarios, in particular the acceleration around massive star wind-termination shocks, were shown to barely accelerate up to PeV \citep{casse1980,cesarsky1983b}, and to suffer several issues such as the problem of the injection and confinement at a perpendicular shock (the magnetic field being azimuthal far away from the massive star). 

%The importance of self-excited magnetic fluctuations was already highlighted in the earliest DSA work \citep{bell1978}, though understanding of this complex plasma process and its role in determining the maximum energy continues to evolve \cite[e.g.][]{MarcowithReview}.
Significant progress followed the demonstration by \citet{Lucek2000}, and in particular \citet{bell2004}, that CRs could drive substantial non-linear amplification of magnetic fields around young SNR shocks.
%Although this mechanism indeed theoretically allows the production of PeV protons, it is nowadays not supported observationnally. In particular, the gamma-ray spectrum of known young SNRs such as Cassiopeia A generically present a cut-off energy well below PeV energies [ref].
This scenario of particle acceleration however requires rather special conditions to reach PeV bands \citep[e.g.][]{ZP2008,bell2013,gabici2019,cristofari2021} and lacks observational confirmation \cite[though see][]{HESSnova2022}.

To date, a convincing scenario for the production of multi-PeV energy protons in our Galaxy is lacking. Such extremely energetic CRs are not only needed to reproduce the CR Galactic-extragalactic transition as measured by ground-based detectors \citep{parizot2014}, but also to account for recent UHE gamma-ray observations \citep[][]{cao2021,cao2021b}. Besides, a number of sources are detected up to hundreds of TeV with no significant cut-off within the detection range of the instrument \citep[e.g.][]{abramowski2012,HESS2016,abeysekara2020,abeysekara2021}. Interestingly, a number of these observations are correlated with massive star clusters or OB associations born within star-forming regions. It was already pointed out in the early 80s that massive stars are often found in clusters and that collective effects between stellar winds and possibly SNRs (referred to as ``SNOBs'' at that time) could play a key role in the acceleration of high energy CRs \citep[e.g.][]{montmerle1979,cesarsky1983}. Although the SNR origin of Galactic CRs has been extensively explored, the peculiarities of the environment within or around a cluster or association have been scarcely considered.

Here we discuss the importance of massive stellar clusters (MSCs) which form within dense molecular clouds. The stellar feedback from the cluster heats the surrounding medium which inflates a cavity around the cluster. The cavity keeps expanding during the whole cluster lifetime. After several Myr, it reaches a size of about 100~pc and is commonly referred to as a superbubble. A number of arguments support the superbubble (SB) origin of CRs 
%(e.g. %\citealt{higdon1998,parizot1999,higdon2006,tatischeff2018,tatischeff2021}, see also the reviews by \citealt{lingenfelter2018,bykov2020}).
\cite[see e.g.][for recent reviews]{lingenfelter2018,bykov2020}.
In particular these environments are expected to excite a very turbulent plasma which effectively confines the particles while a number of shocks sweep up the medium and accelerate the particles, such that a substantial fraction of the stellar power can be efficiently transferred into CRs \citep{montmerle1979,bykov2001}. This was confirmed in the last decade by an increasing number of gamma-ray observations \citep[e.g.][]{Cygnus_FERMI,HESS_Wd1,HESS_LMC,yang2018,aharonian2019,abeysekara2020,mestre2021}. 
This class of sources have generically a spectrum compatible with the local CR flux when propagation effects are taken into account ($\dd N_\gamma/\dd E \propto E^{-2.2 ... -2.5}$). Besides, the composition of the plasma around massive star clusters or associations is expected to differ from that of the standard interstellar medium as it is enriched by the winds of Wolf-Rayet (WR) stars. This may explain a number of composition anomalies measured in the local CR flux \citep[e.g.][]{gupta2020,tatischeff2021}.

The LHAASO observatory has reported the detection of a PeV photon coincident with the Cygnus region \citep{cao2021}, implying CRs of several PeV are produced therein. The maximum energy of CRs produced within MSCs, OB associations or SBs is expected to increase compared to the case of an isolated SNR, because the stellar outflows should in principle be able to excite a strong turbulent magnetic field, which is needed to confine the particles. This follows from the Hillas criterion \citep{hillas1984}, which is commonly invoked to set an upper bound on the maximum energy of the particles accelerated in a given source
%, i.e. that achieved in the electric potential induced by the source, $\Phi = B R u/c$, such that
as $E_{\rm max} < Z e B R u/c$, where $u$, $B$ and $R$ are respectively the typical velocity, magnetic field and size of the source. Assuming $B \approx 10$~\textmu G the magnetic field inside the SB, $R \approx 100$~pc the size of the SB and $U \approx 3000$~km/s the typical velocity of the SNR shocks and stellar winds \citep{bykov1995,bykov2001b}, one obtains $E_{\rm max} \sim 10$~PeV. This simple estimate has long motivated the candidacy of SBs as sources of super PeV protons.

This estimate however relies on the implicit assumption that the particle accelerator is operating on the full scale of 100~pc. This has some support from gamma-ray observations \citep[e.g. in the Cygnus region,][]{abeysekara2021} which display an extended emission over several tens of pc. However, it is also very possible that CRs are accelerated in a subsystem, e.g. a SNR expanding within the SB, and only produce gamma-rays after they have diffused within the cavity and reached the dense and strongly magnetised shell which surrounds the SB. This demonstrates the importance of distinguishing SBs as gamma-ray sources and SBs as particle accelerators. In particular, it is not clear which acceleration mechanism would take place over the whole SB scale, or if there is such a mechanism at all.

The present paper aims at clarifying this question, considering a number of detailed scenarios for the acceleration of particles within SB environments (including compact MSCs or looser stellar associations). The Hillas limitation is justified by theoretical arguments which motivate credible values for the relevant source size, velocity, and magnetic field. In Section~\ref{sec:estimates} we anticipate the results using simple estimates in order to get benchmark values for the maximum proton energies. In Section~\ref{sec:SB_prop} we detail the properties of the SB, in particular how its size, pressure, density, mean magnetic field and mean turbulence velocity evolve. Using physically motivated acceleration mechanisms, we then compute directly the maximum energy considering three candidate accelerators: the SB forward shock (Section~\ref{sec:forwardshock}), the supernovae and wind termination shocks (Section~\ref{sec:primaryshocks}), and the turbulence (Section~\ref{sec:turbulence}). In each case, we discuss the possibility of accelerating UHE (about 10 PeV) protons. We conclude in Section~\ref{sec:conclusions}.

\section{Benchmark estimates}\label{sec:estimates}

%%%%%%%%%%%%%%%%%%%%%%%%%%%%%%%%%%%%%%%%%%%%%%%%%%%%%%%%%%%%%%%%%%%
\begin{table*}
\centering
\begin{tabular}{lcccccc}
\hline
Acceleration mechanism & $U$ [km/s] & $B$ [\textmu G] & $R$ [pc] & $E_{\rm max}$, canonical [PeV] & $E_{\rm max}$, optimistic [PeV] \\ \hline
SB forward shock & 30 & $1 - 10$ & $50 - 100$ & 0.01 & 0.1\\
SNR inside SB & 3000 & $10 - 50$ & $10 - 30$ & 1 & $5$  \\
WTS around a compact cluster & 2000 & $10 - 50$ & $5 - 30$ & 1 & $5$  \\
SNR embedded in a WTS & 5000 & $10 - 50$ & $5 - 30$ & 5 & $10$  \\
HD turbulence & 100 & $1 - 10$ & $50 - 100$ & 0.5 & $1$ \\
Collection of individual winds (loose cluster) & $10 - 100$ & $10 - 50$ & $1 - 10$ & 0.05 & $0.5$ \\
%Magnetar reacceleration & $10^4 - 10^5$ & $10^5 - 10^6$ & $10^{-5} - 10^{-4}$ & 10 & $30$ \\
\hline
\end{tabular}
\caption{Benchmark estimates of the maximum proton energy achievable via various acceleration mechanisms expected to take place inside superbubbles. The canonical values correspond to the values commonly inferred from observations of typical examples, e.g. Westerlund 1 as a young compact cluster, Cygnus OB2 as a loose cluster, etc. Optimistic values choose favourable, yet plausible combinations of the physical conditions.}
\label{tab:benchmark_estimates} 
\end{table*}
%%%%%%%%%%%%%%%%%%%%%%%%%%%%%%%%%%%%%%%%%%%%%%%%%%%%%%%%%%%%%%%%%%%

The Hillas criterion is sometimes formulated as a confinement criterion: $E_{\rm max} \approx Z e B R$, i.e. particles with Larmor radius larger than the size of the accelerator escape. This is the least stringent constraint one can put on the maximum energy and it does not give any clue about the acceleration mechanism. As particles are accelerated in electric fields (or equivalently the motion of scattering centres), considering a specific acceleration mechanism introduces a factor $u/c$ in the above estimate, where $u$ is some characteristic velocity specific to the acceleration mechanism under investigation. The maximum energy is then that achieved in the electric potential $\Phi = B R u/c$, such that $E_{\rm max} \approx Z e B R u/c$. It should be noted that this should still be understood as an upper bound, as other processes, such as losses, may decrease the maximum achievable energy.

While the confinement criterion is rather universal, introducing the ``characteristic velocity of the source'' $u$ is more challenging as it is not always a well-defined quantity. In order to estimate its value, one needs to specify the mechanism driving the acceleration of particles. In this introductory section, we consider a number of possible mechanisms which may work within SBs, in order to derive preliminary limitations on the maximum energy. Each mechanism will be considered in more detail in the following sections.

SBs are complex environments. The hot rarefied plasma is spanned by primary shocks, which decay into turbulent motions and MHD waves. Collective effects such as re-acceleration processes contribute to the acceleration of particles. SBs are also delimited by a forward shock which expands into the ISM. While the size of this shock can be very large ($R \gtrsim 100$~pc, \citealt{weaver1977}), it is too slow ($u \sim 30$~km/s) to accelerate PeV protons, even assuming large magnetic fields in the ISM.
More generally, in the case of strong shocks, the velocity $u$ is identified as the velocity of the shock, as it is precisely the velocity jump at the shock discontinuity which drives the acceleration of the particles via the DSA mechanism \cite[e.g.][]{drury1983}. Inside SBs, there are two types of strong primary shocks: the time-dependent SNR shocks which expand after a supernova (SN) explosion, and the wind termination shocks (WTS) which surround the individual massive stars, or the entire stellar cluster if it is compact enough. SNRs and WTSs have a typical velocity of several $1000$ km/s. In the low-density SB interior, SNRs expand to a typical radius of a few tens of pc before reaching the Sedov-Taylor phase, which is generally larger than the radius of the WTS, even in the case of a WTS powered by a very massive compact cluster. The size of the latter shock depends on the mechanical power of the stellar cluster and is typically of the order of 10~pc \citep{weaver1977}. It is believed that, due to in situ CR acceleration, efficient magnetic field amplification takes place upstream of SNR shocks, leading to a magnetic field of up to several tens of \textmu G. On the other hand, it is less clear if the magnetic field can be as efficiently amplified in stellar winds. In both cases, the maximum energy is generally inferred to be only $\sim 1$~PeV \citep[e.g.][]{gupta2020,morlino2021,vieu2022}. While this is slightly larger than the maximum energy achieved at isolated SNR, atypical conditions would be required for protons to be accelerated well beyond PeV by primary shocks embedded in SBs. A promising situation combining the advantages of both WTS and SNRs might be that of a SNR expanding in a wind profile close to a compact cluster. A powerful cluster may indeed convert a substantial amount of its mechanical energy into turbulence, amplifying magnetic fields up to hundreds of \textmu G in its vicinity, such that particles could be accelerated up to 10~PeV by a powerful SNR propagating in the wind, even in the absence of additional magnetic-field amplification. Considering a similar scenario of SNR shock -- cluster wind interaction with efficient turbulence generation, \citet{bykov2015} found that proton energies up to 40~PeV could be reached in the case of fast shocks ($U = 10^4$~km/s).

The winds and SNRs collectively sustain turbulence in the SB.
Supersonic hydrodynamic turbulence, composed of random large-scale motions of the plasma, provides compression and rarefaction waves which may also be able to accelerate particles. Indeed, the random waves are scattering centres and the original stochastic Fermi acceleration mechanism takes place \citep{skilling1975III,ptuskin1988,bykov1990a}. In this case, the velocity $u$ should be identified as the mean velocity of the flows, that is, the square root of the energy stored in the turbulence. This provides typically $u \sim 100$~km/s, roughly an order of magnitude smaller than the velocity of primary shocks. Even if, in this case, the turbulence is distributed inside the whole SB, such that the relevant size of the accelerator is of the order of 100~pc, the expected maximum energy is estimated as a few hundreds of TeV. If a substantial fraction of the mechanical power of the cluster goes into magnetic field amplification, 1~PeV could be reached, although this is already an optimistic value.

Finally, an ensemble of individual winds, gathered in a \textit{loose} cluster, that is, a cluster around which no collective WTS forms, may be viewed as an ensemble of stochastic shocks from the point of view of high energy particles and act as scattering centres, leading to a stochastic acceleration process over strong primary shocks \citep{bykov2001}. Although the velocity of the scattering centres is of the order of the wind velocity, the scattering cross-section is rather small, i.e. particles diffusing inside the SB have a low probability of encountering an individual WTS multiple times. Indeed, the size of an individual WTS around a massive star is about 1~pc, that is at least several times smaller than the radius of a loose stellar cluster. The average plasma velocity in the stellar cluster is correspondingly about $(1~\text{pc}/R_c)^3 V_w$, where $R_c$ is the radius of the cluster and $V_w \sim 2000$~km/s the average wind velocity.
In the case of an extended cluster, $R_c \gtrsim 5$~pc and the average velocity is of the order of a few tens km/s. The maximum energy achieved by protons via this acceleration mechanism is not expected to exceed 500~TeV. In a linear treatment of the problem, and assuming optimistic parameters ($V_w=3000$~km/s, $B=30$~\textmu G, over a region of radius $R=5$~pc), \citet{klepach2000} showed that PeV energies could indeed be hardly reached via this mechanism of stochastic acceleration. It is furthermore difficult to justify how an average velocity of 3000~km/s could be maintained within an ensemble of stars surrounded by individual wind termination shocks, each of them having a radius much smaller than 5~pc.

Table~\ref{tab:benchmark_estimates} summarises the general expectations on the maximum energy which have been discussed in the present Section. The remainder of the paper is devoted to a more detailed analysis of the possible acceleration mechanisms in the peculiar SB environment.

\section{Superbubble properties}\label{sec:SB_prop}

To motivate the calculations that follow, we introduce the key assumptions of SB environments.
We consider a massive stellar cluster which has formed inside a dense molecular cloud.
Massive stars (which are defined in the following as the stars with mass $M>8 M_\odot$), emit powerful winds during their lives. If the cluster is compact enough, a collective WTS will form around the MSC. Otherwise, each individual star is surrounded by its own WTS.
Massive stars explode as SNe at the end of their lives and turn into expanding SNRs. SNRs and WTS heat the surrounding medium, which inflates a low-density cavity called a superbubble (SB). Thermodynamical considerations provide the expansion law of the SB as \citep{castor1975,weaver1977}:
\begin{align}
\label{eq:weaver}
R(t)
&\approx 68~ \left( \frac{\xi_b \mathcal{P}}{10^{38}~{\rm erg/s}}\right)^{1/5} \left( \frac{n_0}{\rm cm^{-3}} \right)^{-1/5} \left( \frac{t}{\rm Myr} \right)^{3/5} \rm pc
\, ,
\end{align}
where $\mathcal{P}$ is the average mechanical power of the stellar cluster, $n_0$ is the hydrogen number density of the surrounding medium, which typically ranges from 1~cm$^{-3}$ (the average ISM) to 100~cm$^{-3}$ (giant molecular clouds), $t$ is the age of the MSC and $\xi_b \sim 0.1$ a power conversion factor which accounts for the fact that not all the stellar power is converted into superbubble shell expansion \citep[e.g.][]{yadav2017,vieu2022}. In deriving the expression above, it was assumed that $R$ is much larger than the radius of the star cluster and that SNRs become subsonic before reaching the edge of the SB. The average mechanical power of the MSC is roughly constant in time and can be written as $\mathcal{P} = 10^{36} N_*$~erg/s, where $N_*$ is the number of massive stars ($M > 8 M_\odot$) in the MSC \citep{vieu2022}. This includes the main sequence (MS), Wolf-Rayet (WR) and supernova remnant (SNR) phases of the stellar evolution (the slow winds of the red supergiant phase have a negligible contribution \citealt{seo2018}).

The MSC not only blows the SB, it also drives the generation of large-scale MHD turbulence via the decay of primary shocks \citep[e.g.][]{bykov1987,padoan2016} or various instabilities operating at a macroscopic scale \citep[e.g.][]{inoue2009}, along with CR-driven instabilities \citep[e.g.][and references therein]{marcowith2016}. Without delving into details of the turbulence injection mechanism, one can write a simple phenomenological equation for the generation of turbulence as \citep{zhou1990}:
\begin{equation}\label{turbulence_cascade}
\d_t W + \d_k F = S \delta (k-k_0) \, ,
\end{equation}
where $W(k)$ is the turbulent spectral energy density at wavenumber $k$, $F$ is the spectral energy flux from the large scales to the small scales and $S$ the source of the turbulence at the largest scale $2\upi /k_0$. Assuming that the magnetic and hydrodynamic energies are in equipartition, the turbulence spectrum can be normalised such that $\int \dd k W(k) = B^2/(4 \upi) = \rho \bar{u}^2 $, where $B$ and $\bar{u}$ are respectively the turbulent magnetic field and turbulent velocity averaged over the perturbations (or, equivalently, over the volume in the case of homogeneous turbulence). Assuming that the stellar feedback is homogeneous over a volume $V$, one can write the source term in this volume as \citep{vieu2022}:
\begin{align}\label{CRinSB_Sourceturbulence}
    S &= \frac{\eta_T}{V} \mathcal{P} \, ,
\end{align}
where $\eta_T < 1$ is the efficiency of turbulence generation. Numerical simulations suggest $\eta_T \lesssim 30$\% \citep{gallegosgarcia2020}. Further numerical work is needed to better constrain this parameter. In the following numerical estimates, we will assume a reasonable value $\eta_T = 10$\%.

Solving Eq.~\ref{turbulence_cascade} provides typically:
\begin{equation}\label{CR_in_SBturbulenceenergy}
    \frac{B^2}{8 \upi} \approx \rho \frac{\bar{u}^2}{2} \approx \( \frac{\sqrt{\rho} S}{k_0} \)^{2/3} \, ,
\end{equation}
where $\rho$ is the average density in the volume $V$.
Depending on the nature of the turbulence (i.e. the expression of the flux $F$ in Eq.~\ref{turbulence_cascade}), this estimate varies by a small numerical factor of order unity.

The average density and temperature in the SB interior are mainly driven by the evaporation and mass loading at the shell \citep{weaver1977}. Thermodynamical calculations provide \citep[][]{maclow1988}:
\begin{align}
    n(t) &= 0.34  \( \frac{\xi_b N_*}{100} \)^{\frac{6}{35}} \( \frac{n_0}{100~ {\rm cm}^{-3}} \)^{\frac{19}{35}} \( \frac{t}{{\rm Myr}}\)^{-\frac{22}{35}} ~{\rm cm}^{-3} \label{densitySB} \, ,
    \\
    T(t) &= 4.8 \times 10^6  \( \frac{\xi_b N_*}{100} \)^{\frac{8}{35}} \( \frac{n_0}{100~ {\rm cm}^{-3}} \)^{\frac{2}{35}} \( \frac{t}{{\rm Myr}}\)^{-\frac{6}{35}} ~{\rm K} \label{tempSB}
    \, .
\end{align}
The scalings derived in this section describe the thermodynamic, kinetic and magnetic properties of the SB, which we will now use to estimate the maximum energy of the protons accelerated in SB cavities, considering a number of possible scenarios.

\section{Particle acceleration by the superbubble forward shock}\label{sec:forwardshock}
\begin{figure}
          \centering
              \includegraphics[width=\linewidth]{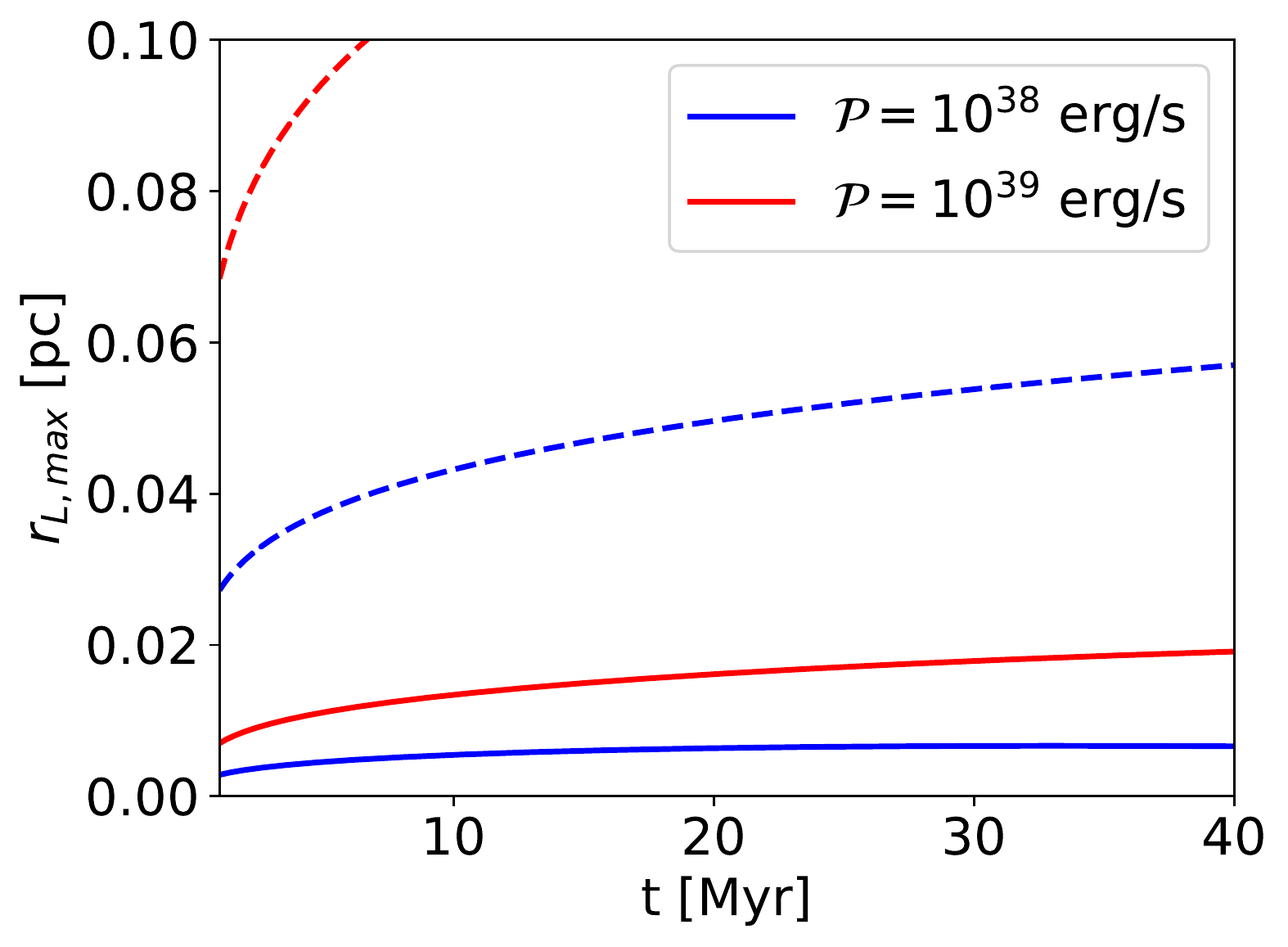}
   \caption{Evolution of the maximum Larmor radius achieved in the SB forward shock around a canonical cluster ($\mathcal{P} = 10^{38}$~erg/s, blue curve) and around a very massive cluster ($\mathcal{P} = 10^{39}$~erg/s, red curve). The solid curves show the limitation due to the finite age of the SB while the dashed curves show the limitation due to the finite size of the SB. We assumed optimistic parameters: $\xi_b=1$, $n_0 = 1$~cm$^{-3}$, and an ISM temperature of $10^4$~K.}
   \label{fig:pmaxfromSBshock}
 \end{figure}

SBs are delimited by a forward shock, whose size and velocity are readily given by Eq.~\ref{eq:weaver}. Even assuming optimistic values for the mechanical power ($\mathcal{P} = 10^{39}$~erg/s) and ISM density ($n_0 = 1$~cm$^{-3}$), the forward shock rapidly becomes weak, e.g. $\dot{R} \approx 65$~km/s at $t=1$~Myr. In the case of a weak shock, the maximum energy achieved by the particles via DSA is generally limited by the available acceleration time.
Assuming that the diffusion coefficient is homogeneous around the SB shock, the acceleration rate reads \citep{lagage1983}:
\begin{equation}\label{accrate}
        \frac{\dd p}{\dd t} \approx \frac{\dot{R}^2 p}{\kappa} \frac{r-1}{3r (r+1)} \, ,
\end{equation}
where $r = (\gamma+1) /((\gamma-1)+2\mathcal{M}^{-2})$ is the shock compression ratio, with $\mathcal{M}$ the shock sonic Mach number and $\gamma$ the adiabatic index of the medium. In order to find the maximum momentum, we assume Bohm diffusion, $\kappa = pc/(\xi qB)$, with $\xi \leq 3$, and integrate the acceleration rate. The Larmor radius of particles at the maximum energy (or rigidity) is plotted in Figure~\ref{fig:pmaxfromSBshock} (solid curves) as function of the age of the SB for optimistic values of the parameters (see caption).

For completeness we also plot on Figure~\ref{fig:pmaxfromSBshock} the size limitation on the maximum energy. Indeed, particles whose diffusion length is larger than the shock radius are expected to escape upstream of the shock. Assuming Bohm diffusion, this provides another limitation as $r_{L,\rm max} = \xi R \dot{R}/c$. This limitation is generically less stringent than the time limitation discussed above.

%%%%%%%%%%%%%%%%%%%%%%%%%%%%%%%%%%%%%%%%%%%%%%%%%%%%%%%%%%%%%%%%%%%
%%%%%%%%%%%%%%%%%%******************************%%%%%%%%%%%%%%%%%%%
%%%%%%%%%%%%%%%%%%%%******************************%%%%%%%%%%%%%%%%%
%%%%%%%%%%%%%%%%%%%%%%%%%%%%%%%%%%%%%%%%%%%%%%%%%%%%%%%%%%%%%%%%%%%
Even in the case of a very massive cluster ($\mathcal{P} = 10^{39}$~erg/s - about 1000 massive stars), the maximum Larmor radius never exceeds 0.02~pc. With more reasonable parameters, e.g. $n_0 \gtrsim 10$~cm$^{-3}$, it falls below 0.01~pc.

In order to estimate the maximum energy from the maximum Larmor radius, one has to specify the magnetic field.
If the magnetic field experienced by the particles diffusing around the forward shock is self-generated via non-resonant streaming instability, one expects $B \propto \rho^{1/2} \dot{R}^{3/2}$ and therefore $B$ decreases with time \citep{bell2004}. In fact the self-generated magnetic field is already negligible at early times as the shock is too slow ($B\approx 0.05$~\textmu G for $N_*=1000$ and $n_0 = 1$~cm$^{-3}$). The fact that after a few Myr the radius of the forward shock becomes of the order of 100~pc is irrelevant in this scenario.

This shows that particles diffusing upstream of the forward shock likely experience the ISM background magnetic field rather than the self-generated magnetic field, and thus the Bohm assumption is already optimistic. The ISM magnetic field is typically a few~\textmu G such that even in the most optimistic case displayed in Figure~\ref{fig:pmaxfromSBshock}, the particles are barely accelerated up to 100~TeV. The ISM magnetic field may be higher in peculiar regions of our Galaxy, such as the Galactic centre, but the ambient density is then also higher than 1~cm$^{-3}$, which slows down the SB expansion. The SB forward shock is therefore generically unable to accelerate PeV protons above 100 TeV, because it rapidly becomes very weak.

\section{Particle acceleration by internal primary shocks}\label{sec:primaryshocks}
As the large-scale collective shock surrounding the SB is unable to accelerate protons to ultra high energies, we now consider the acceleration at internal shocks, which are smaller but faster. Two cases must be distinguished. We first discuss the case of a ``loose'' cluster, where the separation between the clustered stars is too large for a collective outflow to form. In this case, the individual WTS surrounding each massive stars are too small to accelerate very high energy CRs and the only relevant primary shocks are those associated to SNRs. We then discuss the case of a ``compact'' cluster, where the stars are confined in a small region, allowing the formation of a large collective WTS. In this case we consider the acceleration of particles around the WTS itself, and then at SNR expanding in the ``free wind'' upstream of the WTS.

\subsection{Supernova remnants in loose clusters}

\begin{figure}
\centering
    \includegraphics[width=\linewidth]{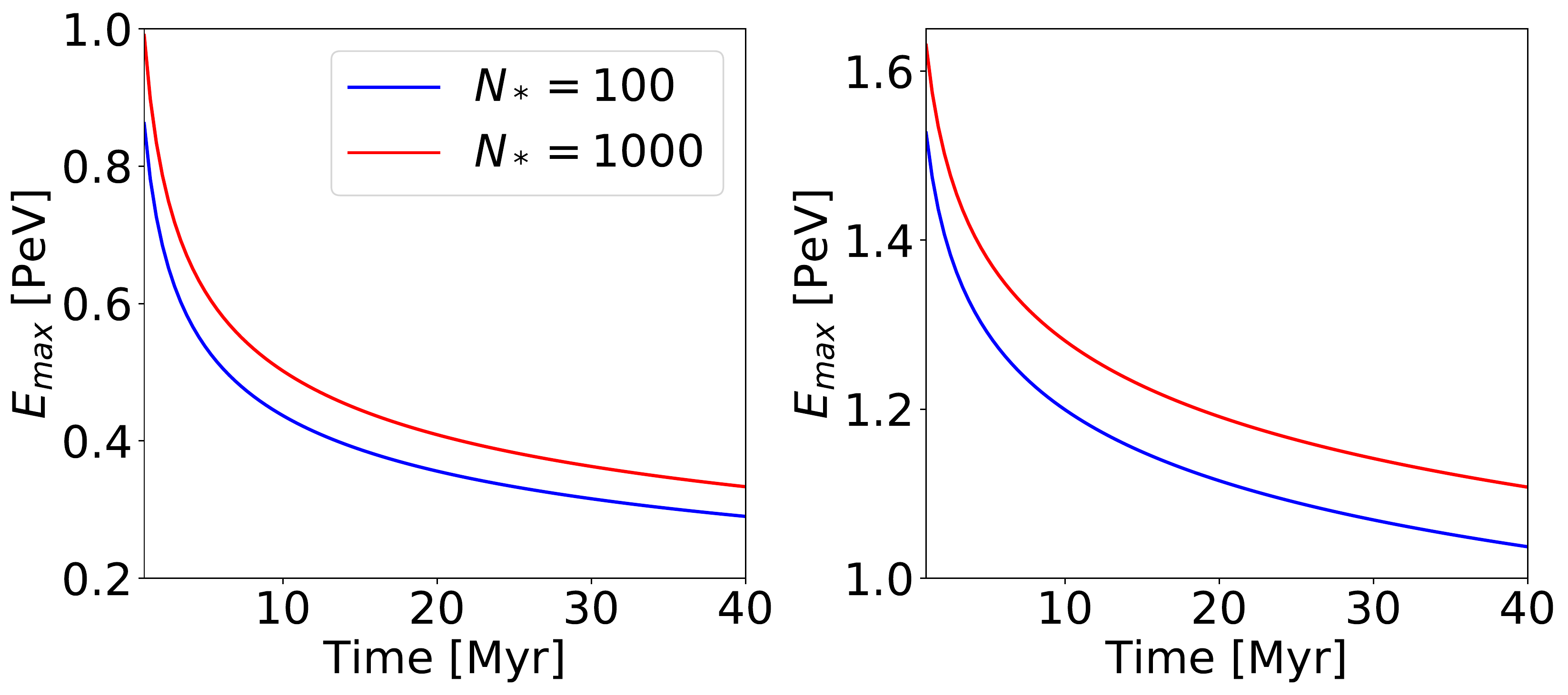}
   \caption{Evolution of the maximum momentum achieved in a SNR. The left panel shows the case where the magnetic field is generated by the stars assuming $n_0 = 1$~cm$^{-3}$, $\eta_T = 10 \%$, $L = 10$~pc. The right panel shows the case where the magnetic field is generated by CR-driven instabilities assuming $n_0 = 100$~cm$^{-3}$, $\eta_{CR} = 10\%$, $L = 10$~pc. The assumed parameters are the ones which provide the most optimistic estimate for the maximum energy.}
   \label{fig:pmaxfromSNRinSB}
 \end{figure}

As stated above, the average stellar separation in a loose cluster is too large for a WTS to form. In this case, the SNRs expand in the low-density cavity for times longer than typical in standard ISM conditions \citep{parizot2004}. In the resulting turbulent medium, it is reasonable to assume a Bohm diffusion regime with a uniform diffusion coefficient. Magnetic fields can be efficiently amplified via two mechanisms: the generation of a MHD turbulent cascade from the largest scales where the massive stars inject a fraction of their kinetic power (Eq.~\ref{CR_in_SBturbulenceenergy}), or CR driven  instability due to strong CR currents \citep{bell2004}. Neglecting the latter for the moment, the maximum energy, achieved at the end of the free expansion phase, can be estimated \citep{vieu2022}:
\begin{align}\label{maxp_SNR}
	E_{\rm max} = 0.6 \frac{B}{10~\text{\textmu G}} \(\frac{u}{3000 ~\text{km/s}}\) \( \frac{M_{\rm ej}}{10~M_\odot} \)^{\frac{1}{3}} \( \frac{n}{0.1 {~{\rm cm}^{-3}}} \)^{-\frac{1}{3}} \text{~PeV} \, ,
\end{align}
where $M_{\rm ej}$ is the mass of the ejecta. Using the magnetic field and density calculated in Section~\ref{sec:SB_prop}, we plot in the left panel of Figure~\ref{fig:pmaxfromSNRinSB} the evolution of the maximum momentum as a function of the cluster age. We adopt an ISM density of 1~cm$^{-3}$ and turbulence generation efficiency $\eta_T = 10 \%$, as stated above. The maximum momentum decreases with time as the SB expands and the turbulence is diluted over a larger volume. For these generally optimistic parameters, PeV energies are approached, though more typical values will provide $E_{\rm max} \sim 0.3 - 1$~PeV,  similar to what is obtained in isolated SNR shocks.

%%%%%%%%%%%%%%%%%%%%%%%%%%%%%%%%%%%%%%%%%%%%%%%%%%%%%%%%%%%%%%%%%%%
%%%%%%%%%%%%%%%%%%******************************%%%%%%%%%%%%%%%%%%%
%%%%%%%%%%%%%%%%%%%%******************************%%%%%%%%%%%%%%%%%
%%%%%%%%%%%%%%%%%%%%%%%%%%%%%%%%%%%%%%%%%%%%%%%%%%%%%%%%%%%%%%%%%%%

Let us then account for the self-generation of the magnetic field upstream of the SNR shock. Under favourable conditions, this may enhance the maximum energy in the case of isolated SNR. It is indeed expected that a fraction of the shock pressure will be converted in magnetic turbulence due to CR-driven instabilities \citep[e.g.][etc.]{Dorfi,bell2004,beresnyak2009,downes2014,xu2017,zweibel}. We write:
\begin{equation}
\frac{B^2}{8 \pi} \approx \frac{u \eta_{CR}}{c} \rho u^2 \, ,
\end{equation}
where $\eta_{CR} \sim 10$~\% is the acceleration efficiency.
Plugging this into Eq.~\ref{maxp_SNR} provides:
\begin{align}
	E_{\rm max} \approx 4 \(\frac{u}{3000 ~\text{km/s}}\)^{5/2} \( \frac{M_{\rm ej}}{10~M_\odot} \)^{\frac{1}{3}} \eta_{CR}^{1/2} \( \frac{n}{0.1~\rm cm ^{-3}} \)^{1/6}\text{~PeV} \, .
\end{align}
This is plotted in the right panel of Figure~\ref{fig:pmaxfromSNRinSB} for $n_0 = 100$~cm$^{-3}$ and $\eta_{CR} = 10$\%. Interestingly, in this case, $p_{\rm max}$ weakly depends on the density inside the SB, and thus is almost constant throughout the SB lifetime.

This shows that even though the CR-driven instabilities are less efficient in a low-density medium, PeV protons could be generically expected from SNR shocks expanding inside SBs. However the maximum proton energy is not found to reach 10~PeV, even around very fast SNR shocks. Besides, as the cavity expands, the turbulence level decreases: if CR-driven instabilities are not effective, PeV protons are only expected around young clusters ($t\lesssim 10$~Myr).

Note that from the scalings provided in Section~\ref{sec:SB_prop}, one can show that the magnetic field is more efficiently generated from the stellar outflows if the ISM density is low. On the other hand, if the magnetic field is generated via CR-driven instabilities, a high ISM density is more favourable. Therefore, the two mechanisms of magnetic field amplification do not generally apply simultaneously.

\subsection{Wind termination shock (WTS)}
\begin{figure}
\centering
\includegraphics[width=0.8\linewidth]{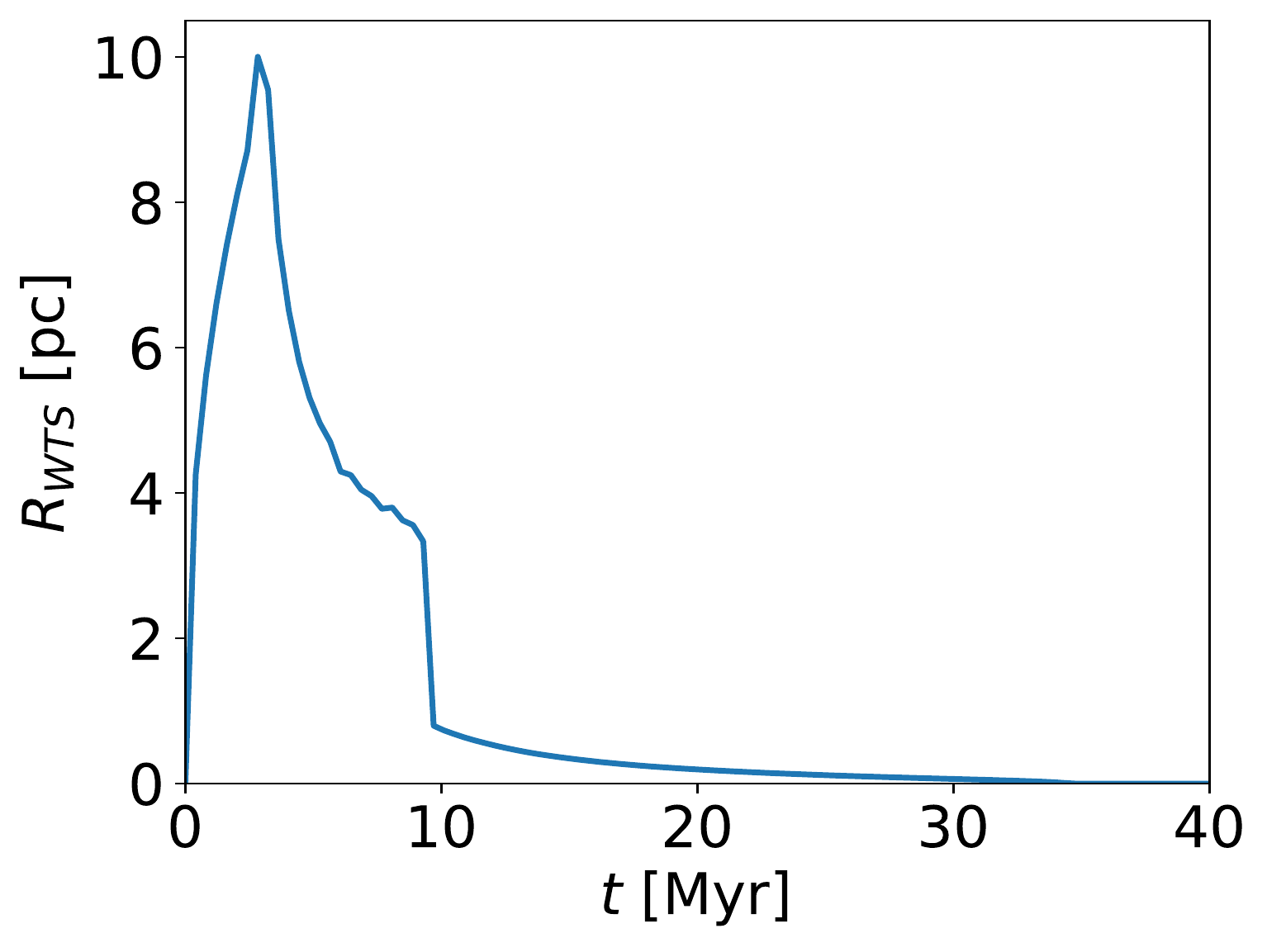}
\caption{Evolution of the radius of the WTS around a cluster of 100 massive stars, statistically averaged over 1000 sample MSCs randomly generated according to a Salpeter initial mass function \citep{salpeter1955}.}
\label{fig:sizeWTS}
\end{figure}

We now consider the case of a compact cluster. Before investigating the properties of the SNRs exploding within compact clusters, we discuss the WTS which surrounds the cluster. Indeed, the outflows expelled by massive stars gathered in a compact cluster pile up onto a collective shock around the cluster. The size of the WTS is determined by the balance between the wind pressure upstream of the WTS and the pressure inside the SB downstream of the WTS. The latter is derived from the average temperature and pressure inside the SB as given by Eqs.~\ref{densitySB} and~\ref{tempSB}. Using the fits of stellar evolution models provided by e.g. \citet[][]{seo2018}, we compute the wind mechanical power, and thus the wind pressure, during the entire lifetime of 1000 random MSC samples generated using a Salpeter initial mass function \citep{salpeter1955} for 100 massive stars. The evolution of the WTS radius averaged over the samples is shown in Figure~\ref{fig:sizeWTS}.
%%%%%%%%%%%%%%%%%%%%%%%%%%%%%%%%%%%%%%%%%%%%%%%%%%%%%%%%%%%%%%%%%%%%
%%%%%%%%%%%%%%%%%%******************************%%%%%%%%%%%%%%%%%%%
%%%%%%%%%%%%%%%%%%%%******************************%%%%%%%%%%%%%%%%%
%%%%%%%%%%%%%%%%%%%%%%%%%%%%%%%%%%%%%%%%%%%%%%%%%%%%%%%%%%%%%%%%%%%

The radius of the WTS reaches a maximum at 3~Myr, then decreases as the most massive stars start to explode, and eventually drops dramatically after 10~Myr, since no WR stars remain in the MSC after this time. At 3~Myr, the wind mechanical power is comparable to the energy deposited by supernova remnants (about $10^{51}$~erg/Myr for a cluster of 100 massive stars). An upper limit on the size of the wind termination shock can be computed from a pressure balance as \citep{vieu2022}:
\begin{equation}\label{CR_in_SB:radiusTS}
R_{WTS} \lesssim 6.2~\xi_b^{-1/5} \(\frac{N_*}{100}\)^{3/10} \( \frac{n_0}{100~{\rm cm}^{-3}} \)^{-3/10} ~{\rm pc}
\, .
\end{equation}
Note that the collective WTS will disappear if its hypothetical radius becomes smaller than the extent of the MSC. This typically happens once all WR stars have exploded, after 10~Myr. The acceleration mechanism discussed in this section is therefore only applicable in general to young and compact MSCs.

The dynamical timescale of the WTS is much larger than the DSA acceleration time. Therefore the maximum energy achievable by particles accelerated at the WTS is limited by the shock size. It was shown by \citet{morlino2021} that the standard Hillas estimate gives an accurate zeroth-order estimate of the maximum energy. The maximum energy is achieved when the diffusion length becomes of the order of the size of the shock, for beyond this limit the particles escape in the upstream region. In the case of Bohm diffusion, $\kappa = pc^2/(\xi qB)$, the maximum momentum is then:
\begin{equation}
E_{\rm max} \approx \xi q B R_{WTS} V_w/c \, ,
\end{equation}
where $\xi = 3r_{\rm g}/\lambda_{\rm mfp}$, with $\lambda_{\rm mfp}$ the mean free path due to magnetic deflections. The most optimistic case, the Bohm limit, corresponds to $\xi=3$.
Even in the scenario where the ISM density is small ($n_0 \sim 1$~cm$^{-3}$) and the star cluster very massive ($N_* \approx 1000$), PeV protons can only be accelerated if the magnetic field is amplified to typically 10~\textmu G. Given that the efficiency of CR-driven instabilities in the wind density profile is still an open question, we conclude that in the absence of other field amplification mechanisms, the proton energies are not expected to reach more than a few PeV, in the most optimistic estimate. This conclusion is in agreement with the in-depth computation of \citet{morlino2021} and the simulations of \citet{gupta2020}.

The magnetic field is also amplified by the generation of large-scale MHD turbulence from the decay of primary outflows emitted by the stars (winds and SNRs within the cluster). As discussed in Section~\ref{sec:SB_prop}, it is expected that a fraction of the mechanical power of the MSC is converted into MHD turbulence within the compact cluster. A major difference with the case of a loose cluster is that the magnetic field is confined. Indeed, the magnetic field advected in the free wind region has a $1/r$ radial profile and the associated energy density is negligible. Let us assume that the stars are evenly distributed within a sphere of radius $R_c$, which is the ``extent'' of the cluster. We then expect the turbulence to be homogeneous in the cluster. The turbulence scale is taken as the distance between the stars, which is $L = (2/3)^{1/3} R_c N_*^{-1/3}$. The magnetic field in the cluster is estimated from Eq.~\ref{CR_in_SBturbulenceenergy} as:
\begin{equation}\label{Bincluster}
    B_c \approx 150 \( \frac{n_c}{10 ~{\rm cm}^{-3}} \)^{\frac{1}{6}}
     \( \frac{\eta_T}{0.1} \)^{1/3} \(\frac{N_*}{100}\)^{2/9} \( \frac{R_c}{1~{\rm pc}} \)^{-2/3} ~\text{\textmu G} \, .
\end{equation}
%Because the magnetic field decreases as $1/r$ in the $1/r^2$ density profile of the free wind, the Alfven velocity is constant upstream of the WTS and reads:
%\begin{equation}
%    v_A \approx 280 \( \frac{n_c}{10 ~{\rm cm}^{-3}} \)^{\frac{-1}{3}}
%     \( \frac{\eta_T}{0.5} \)^{1/3} \(\frac{N_*}{1000}\)^{2/9} \( \frac{R_c}{1~{\rm pc}} \)^{-2/3} ~\text{km/s} \, .
%\end{equation}
Because the magnetic field decreases as $1/r$ outside of the cluster core, the intensity of the magnetic field at the position of the WTS is inversely proportional to the radius of the WTS and the maximum momentum does not depend anymore on the size of the WTS, but rather on the cluster size: $p_{\rm max} \approx 3 q B_c R_c (V_w/c)$ which leads to the following estimate:
\begin{multline}\label{pmaxWTS}
E_{\rm max} \approx 5 \( \frac{\xi}{3} \) \( \frac{n_c}{10 ~{\rm cm}^{-3}} \)^{\frac{1}{6}}
     \( \frac{\eta_T}{0.1} \)^{1/3} \(\frac{N_*}{1000}\)^{2/9} \( \frac{R_c}{1~{\rm pc}} \)^{1/3} \\ \times \frac{V_w}{2000~\text{km/s}} ~{\rm PeV} \, .
\end{multline}
This suggests that WTS around compact clusters can generically accelerate protons up to several PeV. Remarkably, the maximum energy weakly depends on the properties of the cluster, for instance its mass and extension, provided that it is compact and powerful enough to allow the formation of a collective WTS. The numerical values input in Eq.~\ref{pmaxWTS} correspond to those typically assumed for Westerlund 1, where recent HESS observations show a ring-shaped gamma-ray emission enclosing the cluster, which could theoretically correlate with the position of the WTS \citep{mohrmann2021}. To accelerate protons beyond 10~PeV at a MSC WTS requires a very massive ($N_* \sim 1000$, i.e. a total cluster mass of about $5 \times 10^4 M_\odot$) extended ($R_c \sim 3$~pc) cluster, with highly efficient turbulence generation. The only known clusters which displayed such properties are RSGC01, RSGC02, RSGC03 \citep{krumholz2019} but with an age above 10~Myr they are too old to display a WTS: the Wolf-Rayet stars have already exploded several Myrs ago.

\subsection{SNRs in the cluster}
SNRs exploding within young compact clusters are not directly launched in the low-density SB cavity. It is in fact yet unclear what happens to SNR shocks expanding deep inside the cluster. It is very possible that the shock, whose pressure and size are comparable with that of the stellar outflows of the most massive stars, will be disrupted before leaving the cluster. This is expected in particular to enhance the turbulence level and the magnetic fields in the core of the cluster (Eq.~\ref{Bincluster}). Keeping these reservations in mind, let us nevertheless assume that a SNR shock is launched in the free wind region with a relative velocity $V_c$ at $r=R_c$ (e.g. a massive star explodes close to the edge of the cluster). In the $1/r^2$ wind profile, the ejecta-dominated SNR shock follows a self-similar evolution, $R \propto t^m$, with $m \approx 7/8$ \citep[e.g.][]{finke2012,gaggero2018}.
Particles will be accelerated while the shock expands upstream of the WTS \citep{gupta2020}. The maximum energy is limited either by the size or the age of the shock. Because the magnetic field decreases as $1/r$ in the free wind region, the size limitation gives:
\begin{equation}\label{sizelimitSNRcompactcluster}
    E_{\rm max,size} \approx \xi q B_c R_c \dot{R}/c \, ,
\end{equation}
which decreases with time as the shock slows down. On the other hand, the time limitation on the maximum momentum is inferred by considering the acceleration rate (Eq.~\ref{accrate}) in the case of a strong shock ($r=4$). Assuming again Bohm diffusion and integrating on time leads to (for $m<1$):
\begin{equation}\label{timelimitSNRcompactcluster}
    E_{\rm max,time} \approx \frac{\xi}{5} q B_c R_c \frac{V_c}{c} \frac{m}{1-m} \( 1 - \( \frac{R}{R_c} \)^{(m-1)/m} \) \, .
\end{equation}
The maximum energy is achieved when both criteria \ref{sizelimitSNRcompactcluster} and \ref{timelimitSNRcompactcluster} are equal (earlier the SNR is too young, later the SNR is too slow), i.e. when the SNR shock reaches a radius of:
\begin{equation}
    R_* = R_c \( \frac{m}{5-4m} \)^{m/(m-1)} \, .
\end{equation}
However, for $m=7/8$, $R_* = 44 R_c$ which is expected to exceed the radius of the WTS. Therefore the maximum energy will be reached when the SNR collides with the WTS and can be computed by setting $R=R_{WTS}$ in Eq.~\ref{timelimitSNRcompactcluster}. We get, assuming $m=7/8$:
\begin{multline}\label{timelimitSNRcompactcluster2}
    E_{\rm max} \approx 10 \frac{V_c}{5000~\text{km/s}} \( \frac{\xi}{3} \) \( 1 - \( \frac{R_c}{R_{WTS}} \)^{1/7} \) \( \frac{R_c}{1~{\rm pc}} \)^{1/3} \(\frac{N_*}{100}\)^{2/9}
    \\ \times \( \frac{n_c}{10 ~{\rm cm}^{-3}} \)^{\frac{1}{6}}
     \( \frac{\eta_T}{0.1} \)^{1/3}   ~\text{PeV} 
\end{multline}
For reasonable cluster sizes, $R_{WTS}/R_c \sim 5 - 30$, we get $\left(1 - ( R_c/R_{WTS} )^{1/7}\right) \sim 0.2 - 0.4 $ and therefore a maximum energy of the order of several~PeV for the parameters used in Eq.~\ref{timelimitSNRcompactcluster2}. Therefore, if a supernova launches a fast shock (with velocity relative to the wind  $\lesssim 10\,000$~km/s) in the outer parts of a compact cluster which efficiently excites MHD turbulence to generate magnetic fields of about 100~\textmu G in its vicinity, protons could be accelerated around the SNR beyond 10~PeV. Note that the collision between the SNR and the WTS is not expected to enhance the maximum energy \citep{vieu2020}, unless non-trivial amplification effects arise, which is beyond the scope of the present work.

A recent SNR shock expanding in a collective wind around a MSC might be responsible for the acceleration of particles in e.g. Westerlund 1, the Quintuplet, $[DBS2003]~179$. Other very young MSCs such as Westerlund 2, NGC 3603, or the Arches are a priori excluded as no supernova is expected to have exploded within the cluster yet.

\subsection{Reacceleration by SNR shocks}
Around massive star clusters, SNR shocks span the same region successively. The question of particle acceleration by successive time-dependent shocks was investigated in \citet{vieu2021}, where the maximum achievable energy was left as a free parameter. In general, the reacceleration of the particles around an isolated shock is not expected to dramatically enhance the maximum energy compared to the first injection step. Even in the case of interacting winds, efficient reacceleration requires non trivial confinement enhancement \citep{vieu2020}. Eventually, when a SNR launched at the edge of a compact cluster converges toward the collective WTS surrounding the cluster, it will reaccelerate the particles pre-accelerated by the WTS, which may leave detectable spectral features such as bumps or hardenings in the TeV bands \citep{bykov2019,vieu2020}. After the collision between the SNR and the WTS, further particle acceleration may proceed around the transmitted and reflected waves. Although these secondary shocks are expected to be weak
%, with Mach number $M_r = \sqrt{5}$ in the limit of a strong incident shock propagating in a monoatomic gas
\citep{landau1987}, they provide additional reacceleration time which may slightly enhance the maximum energy, by a factor $\sim 2$. 

\section{Particle reacceleration in turbulence}\label{sec:turbulence}

\subsection{Acceleration by compression and rarefaction waves}\label{sec:HDfermiII}
Inside the SB, the particles probe large-scale compression and rarefaction waves. This gives rise to a stochastic acceleration process (second order Fermi mechanism). In order to infer the maximum momentum achievable via this acceleration mechanism, we must investigate the transport of particles in a strongly turbulent medium.

In a strongly turbulent medium, particles diffuse due to their interactions with MHD waves (referred to as ``small-scale diffusion''), but also due to their advection in the random large-scale flows. 
The characteristic diffusion, acceleration and escape times read:
\begin{align}
\tau_d \equiv \frac{L^2}{6 \chi} \, , \qquad
\tau_a \equiv \frac{p^2}{2 D_p} \, , \qquad
\tau_e \equiv \frac{R^2}{6 \chi}
\, , 
\end{align}
where $L$ is the largest (energy-containing) scale of the turbulence ($L=2 \pi/k_0$), $\chi$ is the spatial diffusion coefficient, $D_p$ is the momentum diffusion coefficient and $R$ is the extent of the turbulent region.
A critical parameter driving the acceleration of particles is the ratio of acceleration over escape time, denoted $\theta \equiv \tau_a/\tau_e$. We have:
\begin{equation}
\theta = \frac{3 \chi p^2}{R^2 D_p} \, .
\end{equation}
%From dimensional analysis we expect at first glance, for relativistic particles, $\chi \sim \bar{U} L/3$, $D_p \sim p^2 \bar{U}/L$, which provides $\theta \sim (L/R)^2 \ll 1$. Interestingly, this ratio is independent on i) the particle energy, which means there is a priori no limitation on the maximum energy (of course the limitation will occur when our assumption that the small-scale spatial diffusion is negligible breaks down, as will be discussed later) ; ii) the characteristic velocity of the HD motions (i.e. the strength of the turbulence).

In the following we seek an expression for $\theta$. In order to do so, we compute the transport parameters and associated timescales from first principles following \citet{bykov1990a,bykov1993}.
The starting point is the transport equation in the diffusion approximation:
\begin{equation}\label{TEbasis}
\d_t N = \nabla \cdot \kappa \nabla N - U \cdot \nabla N + \frac{p}{3} \(\nabla \cdot U \) \d_p N \, , 
\end{equation}
where $U$ is the total velocity field including all perturbations and $\kappa$ is the small-scale diffusion coefficient which only includes the interaction between the particles and the magnetic waves, not the effect of the large-scale flows (in contrast to the total diffusion coefficient $\chi$ defined above). Eq.~\ref{TEbasis} describes the small-scale diffusion over MHD perturbations, the advection in the large-scale plasma flows and the energy change due to the interactions with compression and rarefaction waves. It is not possible to solve directly Eq.~\ref{TEbasis} within a semi-analytical framework. Instead, Eq.~\ref{TEbasis} must be first averaged over the HD fluctuations, in order to obtain a transport equation in terms of the mean velocity $\bar{u}$ and not the fluctuating velocity $U$. This average is not a trivial task. Indeed, when the small-scale diffusion length $\kappa/U$ is smaller than the largest eddy turnover length $L$, the particle distribution function $N$ is expected to strongly fluctuate over small-scales and perturbation theory breaks down (within one eddy turnover length, one cannot write $N = \bar{N} + \delta N$ with $\delta N \ll \bar{N}$). One must then renormalise the transport coefficients in order to account for strong long-wavelength fluctuations. \citet{bykov1990a} provides the result as (for isotropic turbulence):
\begin{equation}
    \d_t \bar{N} = \nabla \cdot \chi \nabla \bar{N} + \frac{1}{p^2} \d_p D_p p^2 \d_p \bar{N} \, .
\end{equation}
The interaction with compression and rarefaction waves leads, on average, to a diffusion in momentum space. Besides, the advection in the random large-scale flows leads, on average, to a spatial diffusion, which is included, along with the small-scale coefficient $\kappa$, in the transport parameter $\chi$. The transport coefficients are calculated as \citep{bykov1990a}:
\begin{align}
\chi &= \kappa + \frac{4 \pi}{3} \Re  \iint \dd k \dd \omega k^2 \[ \frac{2 T + S}{i \omega + \chi k^2} - \frac{2 \chi k^2 S}{(i \omega + \chi k^2)^2} \] \, , \label{chiBykov}
%&= \epsilon + \frac{\tilde{\tau_d}}{4 \pi^3} \int \dd x \dd \bar{\omega} x^4 \frac{\( 2 \tilde{T} + 3 \tilde{S} \) \bar{\omega}^2 + \( 2 \tilde{T} - \tilde{S} \) \tilde{\tau_d}^2 x^4}{(\bar{\omega}^2 + \tilde{\tau_d}^2 x^4)^2} \quad [DOUBLE CHECKED]
\\
D_p &= \frac{4 \pi p^2}{9} \Re  \iint \dd k \dd \omega \frac{k^4 S}{i \omega + \chi k^2}
\, , \label{DpBykov}
\end{align}
where $S$ and $T$ are respectively the longitudinal and transverse wave spectra to be specified. We chose the Fourier convention $f(\v{r},t) = \int \dd \v{k} \dd \omega \tilde{f}\(\v{k},\omega \) e^{i\( \v{k} \cdot \v{r} - \omega t \)}$.
In the following we assume a simple isotropic power-law turbulence spectrum:
\begin{multline}\label{turbulencespectrum}
    S(k,\omega) = T(k,\omega) = \frac{\bar{u}^2 (\nu-1) k_0^{-3}}{8 \pi} \(\frac{k}{k_0} \)^{-\nu-2} \\
    \times \( \delta [\omega - \bar{u} k] + \delta [\omega + \bar{u} k] \) H[k-k_0] \, ,
\end{multline}
normalised such that $4 \pi \int \dd \omega \dd k k^2 S(k,\omega) = \bar{u}^2$. $\nu$ is the spectral index of the turbulence energy spectrum (e.g. $\nu = 5/3$ for a Kolmogorov spectrum, $\nu = 3/2$ for a Iroshnikov-Kraichnan spectrum , $\nu = 2$ in the case of supersonic turbulence), and $H[x]$ is the Heaviside step-function.

%%%%%%%%%%%%%%%%%%%%%%%%%%%%%%%%%%%%%%%%%%%%%%%%%%%%%%%%%%%%%%%%%%%%
%%%%%%%%%%%%%%%%%%******************************%%%%%%%%%%%%%%%%%%%
\begin{figure}
\centering
    \includegraphics[width=\linewidth]{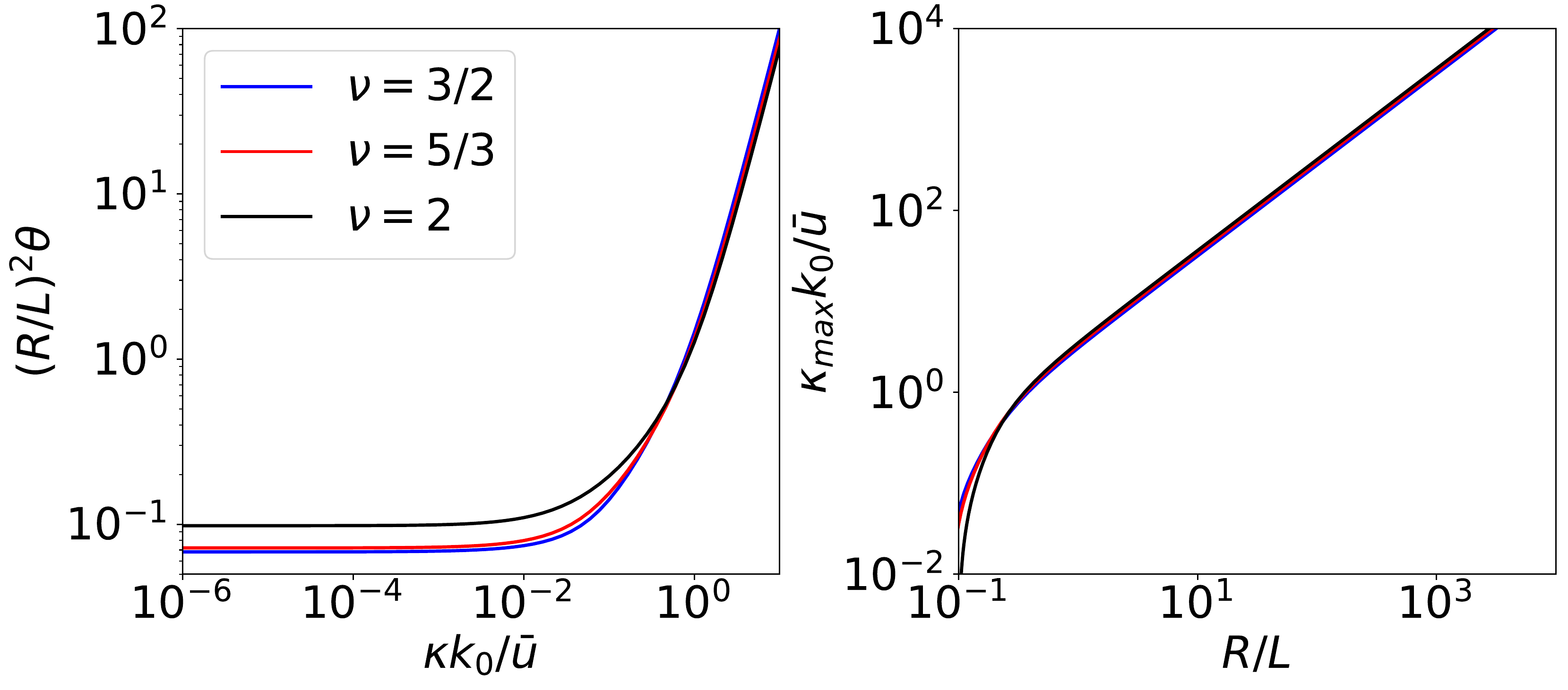}
   \caption{Left: Ratio of acceleration time over escape time as function of the small-scale diffusion coefficient. Right: maximum diffusion coefficient (such that $\theta = 10$) as function of the size of the system.}
   \label{fig:thetakappa}
 \end{figure}
%%%%%%%%%%%%%%%%%%%%******************************%%%%%%%%%%%%%%%%%
%%%%%%%%%%%%%%%%%%%%%%%%%%%%%%%%%%%%%%%%%%%%%%%%%%%%%%%%%%%%%%%%%%%

%Note: if $\tilde{D_p} >\sim \tilde{\chi}$, then one needs to treat the strong acceleration regime, i.e. significant change of momentum within one correlation length (see Bykov 1990). This is not the case here. The ratio $\tilde{D_p}/\tilde{\chi}$ is always smaller than 1 (at most 0.6) such that we can use the weak acceleration approximation.

Using the spectrum~\ref{turbulencespectrum}, one can compute numerically the spatial and momentum diffusion coefficients as given by Eqs.~\ref{chiBykov} and \ref{DpBykov}.
The evolution of the parameter $\theta$ as function of the small-scale diffusion coefficient is shown in Figure~\ref{fig:thetakappa} (left). The transition between the nonlinear regime (transport dominated by large-scale flows requiring the renormalisation procedure) and the perturbative regime occurs for $\kappa \sim 0.1 \bar{u}/k_0$. In the nonlinear regime, $\theta$ is independent on $\kappa$ (and thus on momentum), as the diffusion is driven by the random flows. On the other hand, in the perturbative regime, $\theta$ increases with $\kappa$. The local slope $s$ of the CR spectrum ($f(p) \propto p^{-s}$) can nevertheless be estimated by solving the momentum diffusion equation with constant coefficient, which provides \citep[e.g.][]{bykov2001b}:
\begin{equation}
    s = 3/2+3/2 \( 1+4 \theta/9 \)^{1/2} \, .
\end{equation}
When $\theta$ is momentum-dependent, the bulk of the particle spectrum accelerated via stochastic acceleration in turbulence is not a power-law and the maximum energy is ill-defined. In the following we determine the maximum momentum as that producing $\theta \sim 10$. Above this value the differential energy spectrum becomes steeper than $E^{-3}$, which is %among the steepest power-law spectrum observed in gamma-rays
at the limit of what has been observed in terms of steep gamma-ray spectra
\citep[e.g.][]{cao2021b}.

The plot on the right panel of Figure~\ref{fig:thetakappa} is the same plot as that on the left, but where $\theta = 10$ was fixed, and the axis reversed. This readily shows the value of $\kappa$ above which the acceleration stops, as function of the geometry of the system. If $\kappa$ increases monotonically with momentum, the maximum momentum will always be achieved in the linear regime, as we expect $L < R$ in superbubbles. Similar conclusions hold when one considers more realistic turbulence spectra than the ansatz~(\ref{turbulencespectrum}), e.g. with an exponential decay of the time correlations and a smooth cut-off at $k_0$.

\subsection{Perturbative approach}
The above shows that it is in fact not necessary to implement a nonperturbative procedure to infer a maximum momentum. It is straightforward to show that in the linear regime ($\chi \approx \kappa$) Eq.~\ref{DpBykov} becomes, for any turbulence spectrum:
\begin{equation}\label{linearrelationDpDx}
D_p \kappa \approx \frac{\bar{u}^2 p^2}{9} \, .
\end{equation}
This is a generic result of stochastic particle acceleration in perturbation theory \citep[e.g.][]{ptuskin1988}. A similar result is obtained when considering the stochastic acceleration on MHD waves characterised by a velocity $v_A \approx \bar{u}$ as we assumed equipartition between kinetic and magnetic energy in the turbulence \citep[e.g.][]{thornbury2014}. In this case, the total momentum diffusion, including both the interaction with magnetic and hydrodynamic waves, is twice that given by Eq.~\ref{linearrelationDpDx}. We get:
\begin{equation}
\theta = \frac{27}{2} \( \frac{\kappa}{R \bar{u}} \)^2 \, ,
\end{equation}
and the acceleration stops when $\kappa \approx R \bar{u}$, that is, the expected Hillas limit.

\subsubsection{Diluted turbulence (loose association)}
We now investigate the criterion $\kappa_{\rm max} \sim R \bar{u}$ derived above.
The hydrodynamic turbulence is generated in the SB by the massive stars, with an average flow velocity given by Eq.~\ref{CR_in_SBturbulenceenergy}. The volume of the SB is computed according to Eq.~\ref{eq:weaver}. This procedure implicitely assumes that the stellar energy input is diluted in the whole SB, which is only expected is the stellar cluster is not compact, but rather loose, i.e. no collective wind termination shock forms.

\begin{figure}
          \centering
              \includegraphics[width=0.9\linewidth]{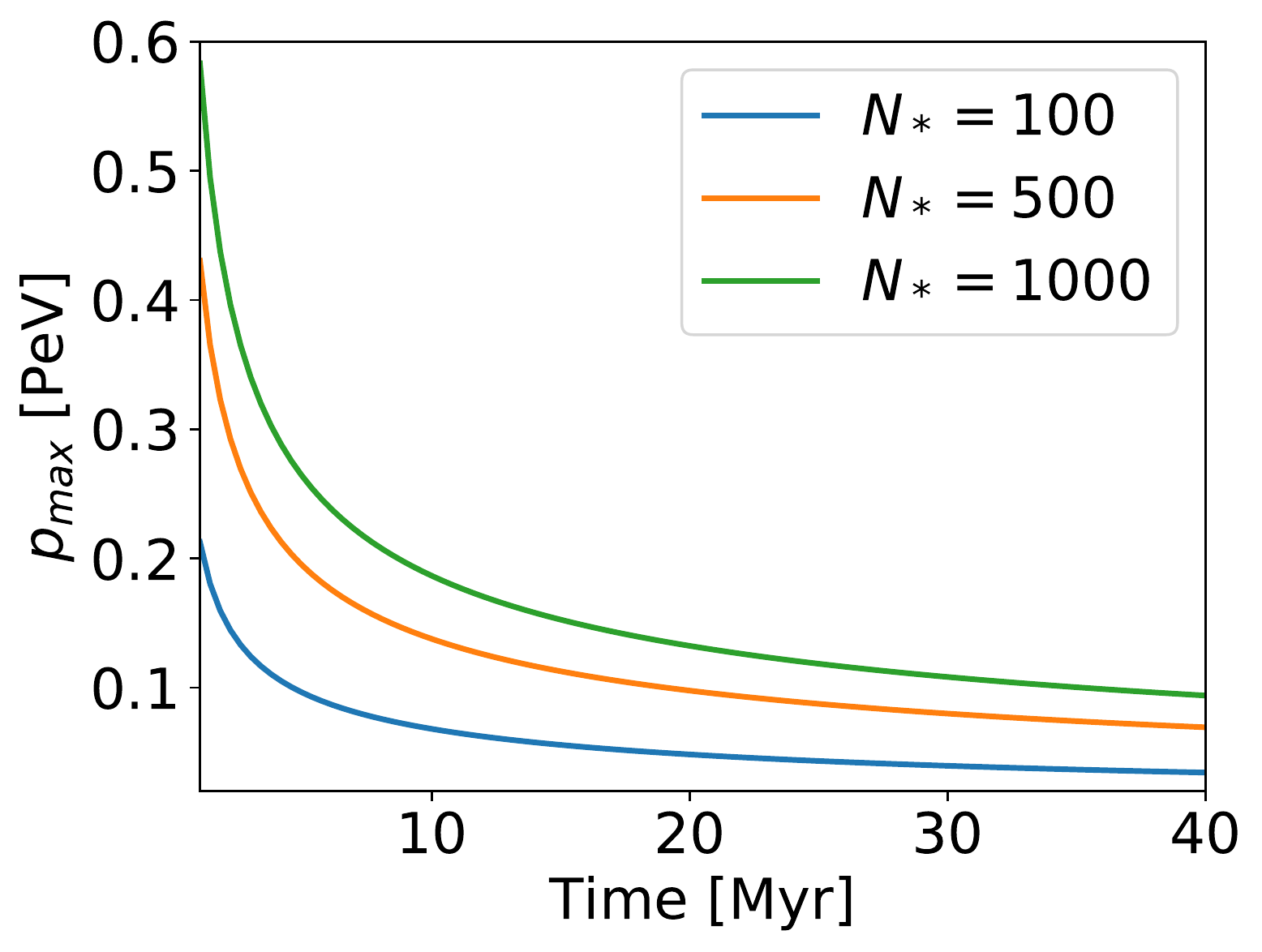}
   \caption{Evolution of the maximum momentum achieved via Fermi II acceleration over diluted turbulence.}
   \label{fig:pmaxFermiII}
 \end{figure}

The estimate of the maximum momentum in the case of small-scale Bohm diffusion ($\kappa \propto p$) is shown in the right panel of Figure~\ref{fig:pmaxFermiII}, as function of the age of the stellar cluster. We took $n_0 = 1$~cm$^{-3}$, $\eta_T = 10$\%, $L= 10$~pc, which are optimistic parameters. Even in this case, only the most massive stellar clusters would produce PeV protons in the few first Myr of their lives. Besides, this result has been obtained assuming a Bohm diffusion regime over the whole SB volume. Harder turbulence spectra such as Kolmogorov or Kraichnan laws are more likely to be generated on this scale \citep[e.g.][]{gallegosgarcia2020}. This would lead to a much larger diffusion coefficient and dramatically decrease the maximum energy of the acceleration mechanism. For instance, it was shown by \citet{ferrand2010} and \citet{vieu2021} that in the case of a Kolmogorov turbulence spectrum, the stochastic acceleration mechanism is irrelevant in SBs beyond GeV bands. Therefore, to accelerate protons to PeV energies and above in a SB filled with a realistic turbulence spectrum would require either an extremely small ambient density, or an unrealistically large turbulent energy density, well above the stellar energy input.

%%%%%%%%%%%%%%%%%%%%%%%%%%%%%%%%%%%%%%%%%%%%%%%%%%%%%%%%%%%%%%%%%%%%
%%%%%%%%%%%%%%%%%%******************************%%%%%%%%%%%%%%%%%%%
%%%%%%%%%%%%%%%%%%%%******************************%%%%%%%%%%%%%%%%%
%%%%%%%%%%%%%%%%%%%%%%%%%%%%%%%%%%%%%%%%%%%%%%%%%%%%%%%%%%%%%%%%%%%

%In the scenario discussed above, there is no distinction made a priori between subsonic turbulence (large-scale random flows) and supersonic turbulence (a collection of weak secondary shocks). This scenario describes how high energy particles can be reaccelerated. High energy particles, whose diffusion length is larger than the largest turbulence scale, do not experience the shock discontinuities, which is why the nature of the turbulence does not matter.

%On the other hand, in order to have particles whose diffusion length is larger than the scale of the turbulence in the first place, a first step of acceleration is required. This first step must be able to accelerate protons up to the energy such that $\kappa =  \bar{U} L$. As the mean turbulent velocity is typically one order of magnitude below the velocity of primary shocks, and the largest turbulence length is at most of the order of the size of the primary shocks, primary shocks, be they SNR or WTS, should be able to provide the first step of acceleration.

\subsubsection{Confined turbulence (compact cluster)}
In the scenario discussed above the turbulence was assumed to be homogeneous over the whole SB, which is only relevant if the stars are evenly spread over a large region. In the case of a compact cluster, the turbulence is efficiently generated within the MSC and rapidly decreases in the cold wind surrounding the cluster. Thus the velocity and magnetic field fluctuations are confined within the MSC. The acceleration region is then the extent of the cluster $R_c$ and not the whole SB. The turbulence scale is taken as the distance between the stars, $L = (2/3)^{1/3} R_c N_*^{-1/3}$. Using Eq~\ref{CR_in_SBturbulenceenergy} we can estimate the properties of the turbulence within the cluster as:
\begin{align}
\bar{u} &\approx 64 \( \frac{n_c}{0.1 ~{\rm cm}^{-3}} \)^{-\frac{1}{3}} \(
     \eta_T N_*^{2/3}\)^{\frac{1}{3}} \( \frac{R_c}{10~{\rm pc}} \)^{-2/3} ~\text{km/s} \, , \label{estimateucluster} 
\\
B &\approx 11 \( \frac{ n_c }{0.1 ~{\rm cm}^{-3}} \)^{\frac{1}{6}} \(
     \eta_T N_*^{2/3}\)^{\frac{1}{3}} \( \frac{R_c}{10~{\rm pc}} \)^{-2/3} ~\text{\textmu G} \, , \label{estimateBcluster} 
\end{align}
where $n_c$ is the density within the boundaries of the cluster.
Then the maximum proton energy reads, assuming Bohm diffusion:
\begin{equation}
    E_{\rm max} \approx 0.18 \, \eta_T^{2/3}
      \(\frac{N_*}{100} \)^{4/9} \( \frac{R_c}{10~{\rm pc}} \)^{-1/3} \( \frac{ n_c }{0.1 ~{\rm cm}^{-3}} \)^{\frac{-1}{6}} ~{\rm PeV} \, .
\end{equation}
Even in the case of a very massive ($N_* = 1000$) and compact ($R_c = 1$~pc) cluster, the maximum energy barely reaches 0.5~PeV in the most optimistic case.

The maximum energy achievable via stochastic acceleration in the turbulent SB is found to be much smaller than previous estimates \citep[e.g.][]{bykov2001b}. In the latter, the region of acceleration was assumed to extend over a size $R_c = 150$~pc and to be filled with strong outflows of mean velocity $\bar{u} = 3000$~km/s. Applying these numbers would imply a maximum proton energy up to 100~PeV.
However, the velocity $\bar{u}$ which is relevant in this acceleration scenario is not that of the primary SNR and wind termination shocks, but that experienced on average by the particles at each scattering. Most of the time, the particles scatter off secondary waves and not primary winds or SNR shocks. In fact, the primary wind outflows contribute to the average outflow velocity in a loose association of $N_*$ massive stars as $\bar{u}_w = V_w N_* (R_s/R_c)^3$, where $V_w$ is the average wind velocity and $R_s$ is the average extent of a single star WTS. In other words, the scattering cross section between the particles and the strong primary shocks is very small. On the other hand, to generate homogeneous turbulence with characteristic velocity of the order of 1000~km/s via the decay of the primary shocks would require an unrealistic stellar power or an extremely low density, as discussed above.

Overall, PeV protons might only be reaccelerated via this mechanism in the Galactic centre region: $\bar{u} \sim 30$~km/s and $B \sim 100$~µG over $R\sim 100$~pc \citep{morris1996} may indeed provide $E_{\rm max} \sim 1$~PeV assuming a Bohm diffusion regime. Such high turbulence level in the Galactic centre is in particular powered by massive star outflows from the Arches, Quintuplet and Nuclear clusters, whose collective feedback might excite strong turbulence and magnetic fields to stochastically accelerate PeV protons.

\section{Conclusions}\label{sec:conclusions}
Young MSCs and SBs have been proposed as promising candidates sources for protons at 1~PeV and above \citep{montmerle1979, aharonian2019, cao2021}. In this work, we have shown that this is in fact not straightforward to accomplish. We considered a number of scenarios with typical environmental parameters, consistently ensuring that the global energy balance is conserved, in particular between the mechanical energy input of the star cluster and the turbulence excited around it. Using physically motivated arguments we have justified that the Hillas limit $\kappa \lesssim \bar{u} R$ can be applied, provided the characteristic velocity of the system $\bar{u}$ is accurately identified, as well as the magnetic field.

Even using optimistic values for the key parameters, the SB forward shock surrounding the whole cavity and the MHD turbulence diluted inside the cavity are generically too weak to accelerate PeV protons. The same conclusion holds when accounting for the contribution of a distribution of strong primary stellar winds, for in this case the scattering cross-section between the particles and the winds is low, even within very massive stellar clusters.

Although primary SNR shocks expanding in a low-density cavity blown by an extended cluster may be able to accelerate protons up to a few PeV if the cluster is young or if CR-driven instabilities operate at their limit, more promising scenarios are found in the case of a compact cluster which powers a collective wind. In this case, the massive stars evolve in a peculiar medium. Indeed, the turbulence is expected to be confined within the cluster as the turbulent energy density released in the collective wind around the cluster is negligible. This implies the generation of large disordered magnetic fields and strong outflows in the cluster core. If the cluster is powerful and compact enough, a large WTS will form during about 10~Myr and accelerate particles. We found that protons up to several PeV are to be expected. Furthermore, SNR shocks may be launched from the edge of the cluster within the free wind region. These will also efficiently accelerate CRs before over-taking the WTS. In this scenario, the most massive compact stellar clusters in our Galaxy such as Westerlund 1 provide channels to efficiently accelerate protons up to about $10$~PeV, without the need for additional self-generated magnetic field amplification. Early computations \citep[e.g.][]{voelk1988} already suggested that SNR shocks expanded in the massive star winds of their progenitors could accelerate PeV protons. The mechanism that we discussed is closely related, though it alleviates the issue of the acceleration efficiency at a perpendicular shock. %Besides, reacceleration effects, in particular around the shock waves reflected after the collision between the SNR shock and the WTS, may further broaden the spectral cut-off, which could be enough to account for the galactic CR population between the ``knee'' and the ``ankle''. This possibility will be investigated in a future work.

Remarkably, our estimate for the maximum energy achieved in this last scenario weakly depends on the mass of the cluster, as long as it is powerful enough to form a collective wind. As most massive stars are expected to be born within compact MSCs, this class of CR sources is expected to be rather common in the Milky Way. Our knowledge of the open cluster population in the Galaxy is still very sparse \citep{cantat2022}. Considerable progress is currently made in the course of the ongoing analysis of the Gaia Data Release 2 and 3, with thousands of open clusters identified and characterised, among which about $\lesssim 100$ are younger than 10~Myr \citep{Cantat2020,tarricq2021}, and therefore expected to power a strongly magnetised collective wind into which SNR shocks will expand. With an average rate of one powerful clustered supernova per 100~kyr per young MSC, we should expect at least one of such event per millennium in the Galaxy, which may be more than enough to account for the CR population in PeV bands \citep{cristofari2020}.

Interestingly, the acceleration of PeV particles around primary shocks embedded in superbubbles always proceeds in a low density environment. Subsequent hadronic interactions are only expected to take place in the dense shell delimiting the superbubble cavity. The incident flux of CR reaching the shell after diffusion in the cavity may generically be too low to produce gamma-ray signatures detectable by the current 100 TeV gamma-ray sensitive instruments such as LHAASO.

%Such promising estimate is only obtained if the magnetic field is efficiently amplified within the compact cluster via the decay of MHD turbulence. Although the energy balance of the cluster allows in principle the generation of large magnetic field (hundreds of \textmu G) within a few pc$^3$, we are lacking MHD simulations to detail the internal properties of very massive compact clusters. It is yet unclear how efficiently the SNRs and stellar outflows convert their mechanical energy into turbulent motion. This will be investigated in future works.

\section*{Acknowledgements}
TV acknowledges S. Gabici, L.M. Bourguinat and L. Härer for helpful discussions.

%%%%%%%%%%%%%%%%%%%%%%%%%%%%%%%%%%%%%%%%%%%%%%%%%%
\section*{Data Availability}
No new data were generated or analysed in support of this research.



\bibliographystyle{mnras}
\bibliography{biblio} % if your bibtex file is called example.bib

%\begin{thebibliography}{99}
%\end{thebibliography}

%%%%%%%%%%%%%%%%%%%%%%%%%%%%%%%%%%%%%%%%%%%%%%%%%%

%%%%%%%%%%%%%%%%% APPENDICES %%%%%%%%%%%%%%%%%%%%%

%\appendix
%\section{Dimensionless integrals}

%%%%%%%%%%%%%%%%%%%%%%%%%%%%%%%%%%%%%%%%%%%%%%%%%%

% Don't change these lines
\bsp	% typesetting comment
\label{lastpage}
\end{document}